\documentclass{ws-mpla}

\usepackage{graphicx}
    \usepackage{latexsym}
    \usepackage{tabularx}
   \usepackage{amsmath}

\begin{document}

\markboth{T. Padmanabhan}
{ Entropy of Horizons, Complex Paths and Quantum Tunnelling}

%
\catchline{}{}{}{}{}
%

\title{Entropy of Horizons, Complex Paths and Quantum Tunnelling
}

\author{\footnotesize T. Padmanabhan}

\address{Inter-University Centre for Astronomy and Astrophysics,\\
Post Bag 4, Ganeshkhind, Pune - 411 007, INDIA.\\
E-mail address: nabhan@iucaa.ernet.in}

\maketitle

\pub{Received (Day Month Year)}{Revised (Day Month Year)}

\begin{abstract}
In any spacetime, it is possible to have a family of observers following a  congruence of timelike curves
such that they do not have access to part of the spacetime. This lack of information 
suggests associating a (congruence dependent) notion of entropy with the horizon that blocks the information 
 from these observers. While the blockage of information is  absolute in classical physics, quantum mechanics will allow 
tunnelling across the horizon. This process can be analysed in a simple, yet general, manner and we show that the probability for a system with energy $E$ to tunnel across the horizon is $P(E)\propto
\exp[-(2\pi/\kappa)E)$ where $\kappa$ is the surface gravity of the horizon.  If the surface gravity changes
due to the leakage of energy through the horizon, then one can associate an entropy $S(M)$ with the horizon
where $dS = [ 2\pi / \kappa (M) ] dM$ and $M$ is the active gravitational mass of the system. The implications are discussed. 

\keywords{entropy, horizons, black hole, complex path, tunnelling, gravitation }
\end{abstract}


Since gravity can affect the trajectories of light rays, it has a strong effect on the light cone structure of the spacetime. The extreme example of this effect occurs when a class of observers [i.e., a congruence of time like curves] find that they cannot receive signals from a region of spacetime, due to the presence of a horizon. This blockage of
information manifests as certain thermodynamic features associated with the horizon in the quantum mechanical context. The purpose of this paper is to provide a remarkably simple way of deriving the temperature of any horizon in a general manner. This is achieved by introducing and utilising the concept of a {\it local Rindler frame}, as an extension of the standard concept of local inertial frame.

Consider an observer moving along a timelike trajectory $X^a(s)$ in a given spacetime. Near any event $\mathcal{P}$ one can choose a locally inertial frame and boost it so that the observer is instantaneously at rest in this frame.
We will call this the local Lorentz frame (LLF). It is usual to assume that purely local measurements of the observer at  $\mathcal{P}$ will coincide with those made in this frame. For example, the quantity $\rho=T_{ab}u^au^b$ is taken to be the energy density since, in the LLF with $u^a=\delta^a_0$, the  $\rho$ reduces to $\rho=T_{00}$.
If we fill a region of spacetime with a congruence of timelike curves $\mathcal{C}$, then this allows us
to define a energy density field $\rho(x)=T_{ab}(x)u^a(x)u^b(x)$ in that region.
\footnote{ Incidentally, the content of Einstein's equations can be
stated entirely in terms of scalar quantities as follows: {\it The scalar projection of $R_{abcd}$ orthogonal to the vector field $u^a$ is $16\pi G\rho$
for all congruences}. The projection is $R_{abcd}h^{ac}h^{bd}=2G_{ac}u^au^c$ where $h^{ac}=(g^{ac}+u^au^c)$
is the projection operator
and the rest follows trivially;
this is a far simpler statement than the one found in some text books \cite{mtw}.}
This relation $\rho(x)=T_{ab}u^au^b$ is remarkable for allowing us to associate a generally covariant scalar field $\rho$
with a non-covariant tensor component $T_{00}(x)$. 
The price we pay is two-fold: (a) The resulting expression $\rho$
is a generally covariant scalar but depends on the congruence $\mathcal{C}$. (b) There is tacit conceptual assumption that one can identify the measurements in LLF with physically meaningful quantities.

In the text book example mentioned above, the velocity field $u^a(x)$ was provided by the congruence. In general, the congruence also provides other vector fields like, for example, the acceleration field $a^i=u^j\nabla_j u^i$ which can be used in a similar manner to construct generally covariant but congruence dependent scalars. In particular,
one can introduce a
 {\it local Rindler frame} at every event in which we can match
 the acceleration of the congruence
 instead of the velocity.
 Since the  acceleration  $a^j=u^i\nabla_iu^j$  is a spacelike vector, it can be mapped to a purely spatial vector $(0, \kappa,0,0)$ at a given event,  which  can be taken to be along the x-axis by a rotation of spatial axes. We now define the   Local Rindler Frame (LRF) near a given event $\mathcal{P}$ by transforming 
from the locally inertial frame coordinates $(T,X,Y,Z)$ around $\mathcal{P}$ to the Rindler coordinates $(t,N,Y,Z)$ through the coordinate transformation:
 \begin{equation}
  \kappa X=N \cosh  \kappa t; \  \kappa T=N \sinh  \kappa t
  \label{keytransform}
  \end{equation}
  The resulting line element near $\mathcal{P}$ is:
   \begin{equation}
   ds^2\simeq -N^2 dt^2 + \frac{dN^2}{\kappa^2} +dL_\perp^2
   \label{dsfirst}
   \end{equation}
   An example of its utility in the classical context is provided by
 derivation of electromagnetic field of an {\it arbitrarily} moving charged particle by coordinate transformations, exactly in the manner in which the
 electromagnetic field of a uniformly moving particle is derived \cite{abhinav}.

Its utility in the study of horizon thermodynamics\cite{Davies} stems from the fact
 the LRF arises very naturally for a 
 general class of metrics with the following properties:
  (i)
 the metric is static  in the given coordinate system, $g_{0\alpha} =0, g_{ab} (t,{\bf x}) = g_{ab}({\bf x})$;
   (ii) the  $g_{00}({\bf x}) \equiv -N^2({\bf x})$ vanishes on some 2-surface $\mathcal{H}$ defined
   by the equation $N^2 =0$, while (iii) $\partial_\alpha N$ is finite and non zero on $\mathcal{H}$
   and (iv) all other metric components and curvature
   remain finite and regular on $\mathcal{H}$.  
 Then the line element will  be:
 \begin{equation}
   ds^2=-N^2 (x^\alpha)  dt^2 +  \gamma_{\alpha\beta} (x^\alpha) dx^\alpha dx^\beta
   \label{startmetric}
   \end{equation}
   The comoving observers in this frame have trajectories ${\bf x}=$
    constant, four-velocity $u_a=-N\delta^0_a$ and four acceleration $a^i=
    u^j\nabla_ju^i=(0,{\bf a})$ which has the purely
    spatial components $a_\alpha=(\partial_\alpha N)/N$.
 The unit normal $n_\alpha$ to the $N=$ constant surface is given by
    $n_\alpha =\partial_\alpha N (g^{\mu\nu}\partial_\mu N  \partial_\nu N)^{-1/2} =
    a_\alpha (a_\beta a^\beta)^{-1/2}$. A simple computation now shows that 
    the normal component of the acceleration $a^i n_i = a^\alpha n_\alpha$, `redshifted'
    by a factor $N$, has the value
\begin{equation}
    N(n_\alpha a^\alpha) = ( g^{\alpha\beta} \partial_\alpha N \partial_\beta N)^{1/2}\equiv Na({\bf x})
\label{defkappa}
\end{equation}
where the last equation defines the function $a$. From our assumptions, it follows that on the horizon $N=0$,
this quantity has a finite limit $Na\to \kappa$; the $\kappa$ is called the surface gravity of the horizon.
These static spacetimes, however,  have a more natural coordinate system defined in terms of the level surfaces of $N$.
 That is, we transform from the original space coordinates $x^\mu$ in Eq.(\ref{startmetric}) to the set $(N,y^A), A=2,3$ by treating $N$ as one of the  spatial coordinates. 
 Near the $N\to 0$ surface, $Na\to \kappa$, the surface gravity, and the metric reduces to the Rindler form in Eq.(\ref{dsfirst}) obtained in the
   LRF near the horizon (for a more detailed discussion see \cite{Padmanabhan:2003gd}).
  
   The bad behaviour of the metric in Eq. (\ref{dsfirst})
  near $N=0$ is a feature and not a bug. It is connected with the fact that the observers at constant-${\bf x}$  perceive a horizon at $N=0$ and related
{\it non-trivial physical phenomena}.  
Classically, this phenomenon is just the one way character of the horizon and is well understood. Quantum field theory, however, brings in new features since it permits processes  which are classically forbidden, allowing, for example, tunnelling across the horizon thereby  endowing the horizon with new properties\cite{Parikh:1999mf}.
We will provide a general description of these phenomena using LRF.

To see how this comes about, let us recall that consistent formulation of quantum field theory requires
  analytic continuation in the time coordinate $t$.
  Consider what happens to the coordinate transformations 
  in Eq.~(\ref{keytransform}) and the metric near the horizon, when the  analytic continuation
 $T\to T_E= T e^{i\pi/2}$ is performed. The 
      hyperbolic trajectory of a an observer stationary in the LRF at $N=1$ (for which $t$ measures the proper time), is given in parametric form as $\kappa T=\sinh\kappa  t, \kappa  X= \cosh \kappa t$. 
This becomes a 
      circle, $\kappa T_E=\sin \kappa  t_E, \kappa X= \cos \kappa t_E$ with, $-\infty<t_E<+\infty$
      on analytically continuing in
     both $T$ and $t$. 
      It is now clear that the \emph{complex t-plane probes the region which is classically inaccessible} to the family of observers on $N=$ constant trajectory\cite{Padmanabhan:2003dc}.
The transformations in Eq.~(\ref{keytransform}) with $N>0, -\infty<t<\infty$ cover \emph{only} the right
    hand wedge [$|X|>|T|, X>0$] of the Lorentzian sector. 
     Nevertheless, \emph{both $X>0$ and $X<0$} are covered by different ranges of the ``angular" coordinate $t_E.$ The range $(-\pi/2) < at_E <(\pi/2)$ covers $X>0$ while the range $(\pi/2) < at_E <(3\pi/2)$
    covers $X<0$ etc.
     These results have two important consequences which we will use:
\begin{itemize}
\item
 The light cones of the inertial frame $X^2=T^2$ are mapped to the origin of the $T_E-X$ plane (The region ``inside" the horizon
    $|T|>|X|$ simply \emph{disappears} in the Euclidean sector. ) making physics near the horizon to be localized near the origin of $(T_E,X)$ plane. {\it It is this fact which allows one to use a  local Rindler frame to capture the 
relevant physics. }   
\item    
     Eq.~(\ref{keytransform}) shows that
     $\kappa t \to\kappa  t-i\pi$ changes $X$ to $-X$, ie., trajectory with complex value for $t$ coordinate can
     describe the tunnelling across the horizon, from  $-X$ to $X$. (Performing this operation twice shows that 
     $\kappa t \to \kappa t - 2i\pi$ is an identity transformation implying periodicity in 
     the imaginary time $i\kappa t =\kappa  t_E$. This fact is used in one way or another in several derivations of the temperature associated with the horizon \cite{Damour:76}.)
\end{itemize}      

   We shall now provide a simple but quite general derivation of the thermodynamic properties of the horizon
   using these facts. Consider a physical system (it could be a particle but our description is fairly general)
    described by 
     a wave function $\Psi(t,N,{\bf X}_\perp; E) = \exp[iA(t,N,{\bf X}_\perp ; E)]$ in the WKB approximation
     where $A$ is the solution of  Hamilton Jacobi equation for the system corresponding to energy $E$.
      The dependence of the quantum mechanical
     probability $P(E) =|\Psi|^2$ on the energy $E$ can be quantified in terms of the derivative
     \begin{equation}
     \frac{\partial \ln P}{\partial E} \approx -\frac{\partial}{\partial E}2(\textrm{Im} A) = 
     -2 \textrm{Im} \left(\frac{\partial A}{\partial E}\right)
     \label{pofe}
     \end{equation}
     in which the dependence on the coordinates is suppressed. 
     Under normal circumstances, action will be real in the leading order approximation
     and the imaginary part will vanish. (One well known counterexample is in the case of tunnelling
     in which the action acquires an imaginary part; Eq.~(\ref{pofe}) correctly describes
     the dependence of tunnelling probability on the energy in the case of radioactive decay, for example.)
     For any Hamiltonian system, one can set $(\partial A/\partial E)=-t_0=$ constant
     thereby determining the trajectory of the system.
     Once the trajectory is known, this equation determines $t_0$ as a function of 
     $E$. Hence we can write 
     \begin{equation}
     \frac{\partial \ln P}{\partial E} \approx 2\textrm{Im} \left[ t_0 (E)           \right]
     \label{imto}
     \end{equation}
      We now only need to note that $t_0(E)$
     can pick up an imaginary part if the trajectory of the system crosses the 
     horizon. In fact, since $\kappa t \to\kappa  t-i\pi$ changes $X$ to $-X$ 
      the imaginary
     part of $t_0$ for trajectories that cross the horizon is given by $(-\pi/\kappa  )$ leading to  $(\partial \ln P/\partial E) = -2\pi/\kappa $.
     Integrating, we find that  the probability for the trajectory of any system to cross the horizon,
     with the energy $E$ will be given by the Boltzmann
     factor 
     \begin{equation}
     P(E) \propto \exp \left[- \frac{2\pi}{\kappa } E\right] = P_0\exp \left[- \beta  E\right] 
     \end{equation}
      with temperature $T=\kappa /2\pi$. (For special cases of this general result see \cite{Keski-Vakkuri:1997xp} and references
      cited therein. The sign of imaginary part in $\kappa t \to\kappa  t-i\pi$ is decided by the sign of $E$ and is chosen such that we are considering the tunnelling of a positive energy particle.)
      
      In obtaining the above result, we have treated $\kappa$ (which is determined by the background geometry)
      as a constant  independent of $E$. A more interesting situation develops if the surface gravity of the horizon changes when some amount of energy crosses it.  Suppose, $\kappa(M)$ denotes the surface gravity when the horizon is generated by an active gravitational mass $M$. When the energy $E$ tunnels through the horizon, the surface gravity becomes $\kappa(M-E)$ and the above result generalises to
      \begin{equation}
     P(E) \propto \exp -\int  \frac{2\pi dE}{\kappa(M-E)} \propto \exp S(E)
     \label{pofefromsofe}
     \end{equation}
 where 
 \begin{equation}
 dS\equiv (2\pi/\kappa(M))dM=dM/T(M)
 \label{s}
 \end{equation}
  suggests a natural definition for an entropy function. An explicit example in which this situation arises is in the case of a spherical shell of energy $E$ escaping from a
 a black hole of mass $M$. This changes the mass of the black hole to $(M-E)$ with the corresponding change in the surface gravity. 
 Clearly,  Eq.~(\ref{s}) gives an entropy that is proportional to the area of the horizon in the spherically symmetric spacetimes. 

 The analysis, so far, did not require and special assumptions and is completely general. We shall now discuss
 the conditions under which one can prove that the entropy is related to the area of the horizon. One possible line of attack is the following: In any static spacetime, it is possible to prove the relation:
\begin{equation}
\int_{\partial\cal V}\sqrt{\sigma}d^2x(Nn_\mu a^\mu)=4\pi G M
\label{defs}
\end{equation}
which relates the flux of surface gravity to the Tolman-Komar  \cite{tolman30} active gravitational mass
\begin{equation}
\label{defe}
M=2\int_{\cal V}d^3x\sqrt{\gamma}N (T_{ab}-\frac{1}{2}Tg_{ab})u^au^b
\end{equation}
contained inside the compact surface. When the surface is a horizon the left hand side of Eq.(\ref{defs}) reduces to
$\kappa A_\perp$ where $A_\perp$ is the horizon area. If some amount of energy tunnels through the horizon, then
all the quantities $\kappa,A_\perp$ and $M$ will change and we get the relation $A_\perp\Delta\kappa+\kappa\Delta A_
=4\pi G\Delta M$. Using this relation and Eq.(\ref{s}) and rearranging the terms, we get the expression for the entropy to be:
\begin{equation}
S=\frac{1}{2}A_\perp+\frac{1}{2}\int A_\perp \frac{dT}{T}
\label{gs}
\end{equation} 
To proceed further we need a relation between the horizon area and horizon temperature so that the integral in the second term can be evaluated. This is difficult to obtain
 since we have no  way of relating the horizon area to horizon temperature in the {\it general}
case. (This issue is under investigation.)
In the simplest case
of spherically symmetric metrics with $-g_{tt}=g^{rr}=f(r)$, the area of horizon will be $4\pi a^2$ where
$a$ is a root of the equation $f(r)=0$ while the temperature is related to $f'(a)$; in general, the function $f(r)$ and its derivative are independent and any relation between them is deeply embedded in the constraint equations of gravity. However, one can show by a detailed investigation (see the first work cited in ref.11) that the correct results are
indeed  obtained in this case. Of course, it is straight forward to check that Eq.(\ref{gs}) gives the correct result for the familiar cases like Schwarszchild, de Sitter etc. In both Schwarszchild and de Sitter spacetimes, the second term in Eq.(\ref{gs})---
which arises from the variation of surface gravity --- contributes $(-1/4)A_\perp$, thereby leading to the correct final result of $S=(1/4)A_\perp$. Clearly, Eq.(\ref{gs}) deserves further study.

Finally, motivated by these considerations, one would like to explore the possibility of associating an entropy with area of the horizon in a more general context\cite{Padmanabhan:2003pk}.  A natural definition for static horizons is provided by:
\begin{equation}
S=\frac{\beta}{8\pi G}\int_{\partial\cal V}\sqrt{\sigma}d^2x(Nn_\mu a^\mu)
\label{defsa}
\end{equation}
The integration is over a three dimensional region ${\cal V}$ with a boundary ${\partial\cal V}$ and $\beta=2\pi/\kappa$ where $\kappa$ is the surface gravity of the horizon.
(This is generally covariant but congruence dependent.)
We take this quantity to be the definition of gravitational entropy for any static spacetime with a horizon, based on the following considerations:
First, if the boundary ${\partial\cal V}$  is a standard black hole horizon, 
$(Nn_\mu a^\mu)$ will tend to a constant surface gravity $\kappa$  and the using $\beta\kappa=2\pi$ we get $S={\cal A}/4G$
where ${\cal A}$ is the area of the horizon. Thus, in the familiar cases, this does reduce to the standard expression
for entropy. Second, if the boundary ${\partial\cal V}$ is a compact surface enclosing a compact horizon ${\cal H}$
and if the
region between ${\partial\cal V}$ and ${\cal H}$ is empty, then again we get the entropy $S={\cal A}/4G$
because the flux through the two surfaces are the same when the in between region has $T_{ab}=0$.
 Similar considerations apply to each piece of {\it any} area element when it acts as a horizon for some Rindler observer. Finally, the results obtained in a series of previous papers \cite{Padmanabhan:2002sh}
 showed that the bulk action for gravity can be obtained from a surface term in the action, if we take the entropy
of any horizon to be proportional to its area with an elemental area $\sqrt{\sigma}d^2x$ contributing an entropy
$dS=(Nn_\mu a^\mu)\sqrt{\sigma}d^2x$. Our definition is the integral expression of the same.

 
 \bibliography{imagaction}
   
 \end{document}